# Engineering $Co_2MnAl_xSi_{1-x}$ Heusler compounds as a model system to correlate spin polarization, intrinsic Gilbert damping and ultrafast demagnetization

C. Guillemard[1,2], W. Zhang[1*], G. Malinowski[1], C. de Melo[1], J. Gorchon[1], S. Petit-Watelot[1], J. Ghanbaja[1], S. Mangin[1], P. Le Fèvre[2], F. Bertran[2], S. Andrieu[1*]

[1] *Institut Jean Lamour, UMR CNRS 7198, Université de Lorraine, 54500 Nancy France*

[2] *Synchrotron SOLEIL-CNRS, Saint-Aubin, 91192 Gif-sur-Yvette, France*

**Abstract:**

Engineering of magnetic materials for developing better spintronic applications relies on the control of two key parameters: the spin polarization and the Gilbert damping responsible for the spin angular momentum dissipation. Both of them are expected to affect the ultrafast magnetization dynamics occurring on the femtosecond time scale. Here, we use engineered $Co_2MnAl_xSi_{1-x}$ Heusler compounds to adjust the degree of spin polarization P from 60 to 100% and investigate how it correlates with the damping. We demonstrate experimentally that the damping decreases when increasing the spin polarization from $1.1 \cdot 10^{-3}$ for $Co_2MnAl$ with 63% spin polarization to an ultra-low value of $4 \cdot 10^{-4}$ for the half-metal magnet $Co_2MnSi$. This allows us investigating the relation between these two parameters and the ultrafast demagnetization time characterizing the loss of magnetization occurring after femtosecond laser pulse excitation. The demagnetization time is observed to be inversely proportional to 1-P and as a consequence to the magnetic damping, which can be attributed to the similarity of the spin angular momentum dissipation processes responsible for these two effects. Altogether, our high quality Heusler compounds allow controlling the band structure and therefore the channel for spin angular momentum dissipation.

* corresponding authors :
wei.zhang@univ-lorraine.fr
stephane.andrieu@univ-lorraine.fr

# I - INTRODUCTION

During the last decades, extensive magnetic materials research has strived to engineer denser, faster and more energy efficient processing and data storage devices. On the one hand, a high spin polarization has been one of the most important ingredients that have been seek [1]. For example, the spin polarization is responsible for a high readout signal in magnetic tunnel junction based devices [2,3]. Additionally, a high spin polarization results in a decrease of the threshold current for magnetization reversal by spin torques [4] required for the development of spin-transfer-torque magnetic random access memory devices [5], for gyrotropic dynamics in spin-torque nano-oscillators [6] and for magnetic domain wall motion [7]. On the other hand, the intrinsic magnetic energy dissipation during magnetization dynamics, which is determined by the Gilbert damping constant, needs to be low in order to build an energy efficient device. Fortunately, spin polarization and damping are usually closely related in magnetic materials.

Nowadays, manipulation of the magnetization on the femtosecond timescale has become an outstanding challenge since the demonstration of subpicosecond magnetization quenching [8] and magnetization reversal on the picosecond timescale [9]. Despite the theoretical and experimental work that has been reported up to now, the relationship between the polarization at the Fermi level or the magnetic damping and the ultrafast demagnetization excited by femtosecond lasers, remains unclear [10-15]. Indeed, numerous mechanisms have been proposed but no consensus has yet been reached. In particular, efforts have been undertaken to unify the magnetization dynamics on the nanosecond timescale and the ultrafast demagnetization considering that the spin-flip mechanisms involved in both phenomena could be the same [10-11,16]. Regarding the influence of the damping on the demagnetization time, different predictions have been reported both experimentally and theoretically. In this situation, the need for engineered samples in which the spin-polarization and magnetic damping are well controlled is of utmost importance to unveil their role on the ultrafast magnetization dynamics.

Heusler compounds are a notable class of magnetic materials allowing for tunable spin-polarization and magnetic damping [17]. The absence of available electronic states in the minority band at the Fermi level leads to very high spin polarization and ultra-low damping due to a strong reduction of spin scattering [18-23]. Recently, ultra-low damping coefficient associated with full spin polarization at the Fermi energy was reported in $Co_2Mn$-based Heusler compounds, [22-23]. Among those alloys, $Co_2MnSi$ has the smallest damping down to $4.1 \times 10^{-4}$ with 100% spin-polarization while $Co_2MnAl$, which is not predicted to be a half-metallic magnet, has a damping of $1.1 \times 10^{-3}$ and a spin-polarization of 60 %.

In the present work, we used $Co_2MnAl_xSi_{1-x}$ quaternary Heusler compounds grown by Molecular Beam Epitaxy (MBE) to tune the spin-polarization at the Fermi energy. Controlling the amount of Al within the alloys allows tuning the spin-polarization from 60 to 100 % as measured by spin resolved photoemission. We show that the magnetic damping parameter for these alloys is among the lowest reported in the literature and decreases when the spin-polarization increases. Ultrafast magnetization dynamics experiments were thus performed on these prototype samples. This complete experimental characterization allows us to directly correlate the ultrafast magnetization dynamics to these parameters and comparing our results to the different theory discussed above.

The $Co_2MnSi$ compound grows in the $L2_1$ structure whereas the $Co_2MnAl$ compound grows in the B2 phase as shown by STEM-HAADF analysis [22]. Such different structures are directly observable during the growth by Reflexion High Energy Electron Diffraction (RHEED) since the surface lattice is different for both compounds. Indeed, half streaks are observed along $Co_2MnSi$ [110] azimuth due to the $L2_1$ chemical ordering [24] which is not the case for $Co_2MnAl$ [22]. The RHEED analysis on $Co_2MnAl_xSi_{1-x}$ films with x=0, ¼ ,½ ,¾ ,1 reveals a regular decrease of these half-streaks intensity with x (Figure 1a). This information that concerns only the surface is confirmed in the entire thickness of the films by using x-ray diffraction. Indeed, the (111) peak typical of the chemical ordering in the $L2_1$ structure clearly decreases and disappears with x (Figure 1b).

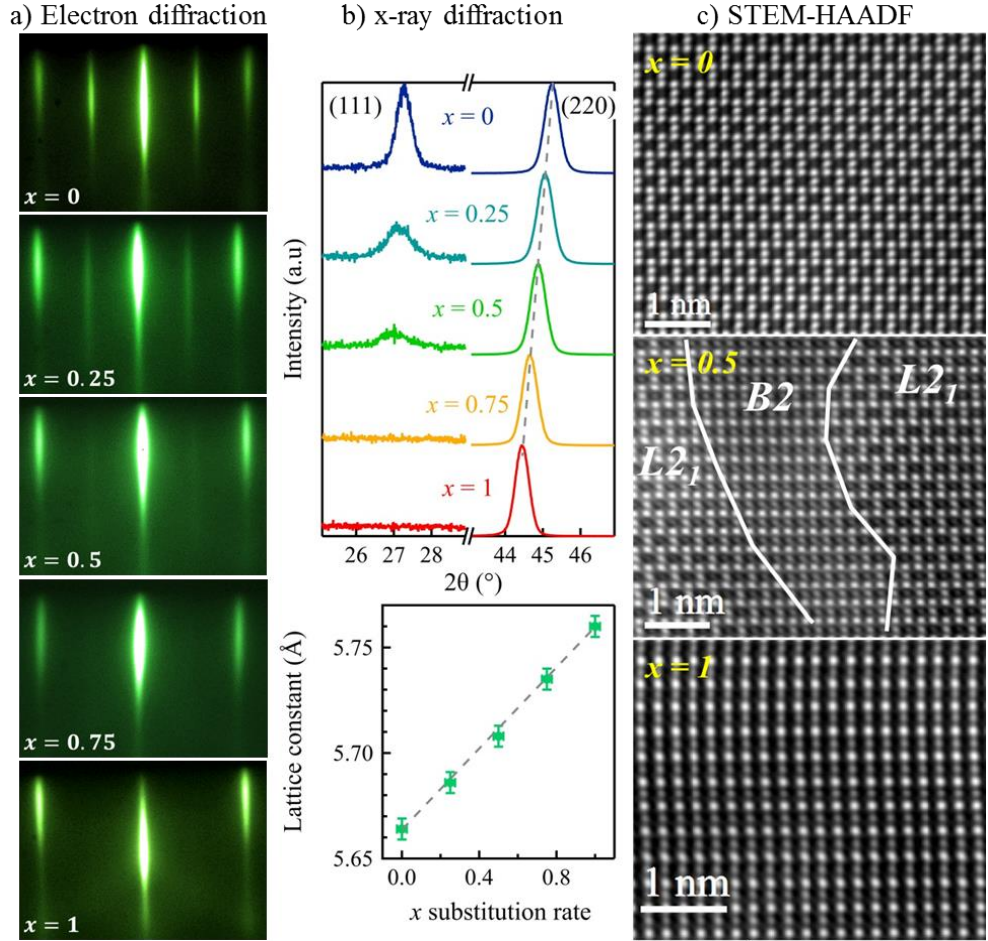

*Figure 1 : a) RHEED patterns along [110] showing the progressive vanishing of the half-streaks (observed on $Co_2MnSi$, x=0) at the surface with x. b) Confirmation of the transition from $L2_1$ to B2 chemical ordering in the entire film by the vanishing of the (111) peak and displacement of (220) peak with x as shown by x-ray diffraction. c) Spatial distribution of both chemical ordering in the films deduced from STEM-HAADF experiments: as the $L2_1$ structure is observed in the entire $Co_2MnSi$ film (x=0), and the B2 one in $Co_2MnAl$ (x=1), a mixing of both structure is clearly observed for x=0.5.*

In addition, the displacement of the (220) peak with x allows us to extract a linear variation of the lattice constant (Figure 1b), as observed in the case of a solid solution. This is an indication that the $L2_1$ chemical ordering progressively vanishes when increasing the Al substitution rate $x$. However, the chemical disorder distribution in the films cannot be easily determined by using the electron and x-ray diffraction analyses. To address this point, a STEM HAADF analysis has been carried on the $Co_2MnAl_{½}Si_{½}$ films with a comparison with $Co_2MnSi$ and $Co_2MnAl$. A clear mixing of both structures is observed for x=½ where around 50% is L2$_1$ chemically ordered and 50%, B2, with typical domains size around 10nm along the growth axis (001) and a few nm in the plane of the film (Figure 1c).

The electronic properties of the Co$_2$MnAl$_x$Si$_{1-x}$(001) series were studied using spin-resolved photoemission (SR-PES) and ferromagnetic resonance (FMR). The SR-PES spectra were obtained by using the largest slit acceptance of the detector (+/- 8°) at an angle of 8° of the normal axis of the surface. Such geometry allows us to analyze all the reciprocal space as confirmed by similar experiments but performed on similar polycrystalline films [23]. Getting the spin-polarization dependence with *x* using raw SR-PES spectra is however not obvious due to the existence of surface states systematically observed on Co$_2$MnSi but also on other Co$_2$Mn-based Heusler compounds [19, 22-23]. To get the bulk spin polarization, we thus used the S polarization of the photon beam. Indeed, we have shown that the surface states are no more detected due to their symmetry [19] without any loss of information on the bulk band structure [23]. The corresponding SR-PES spectra are shown in figure 2. As expected, we thus obtain a tunable spin polarization at $E_F$ from 100% to 63% by substituting Si by Al, as shown in figure 3.

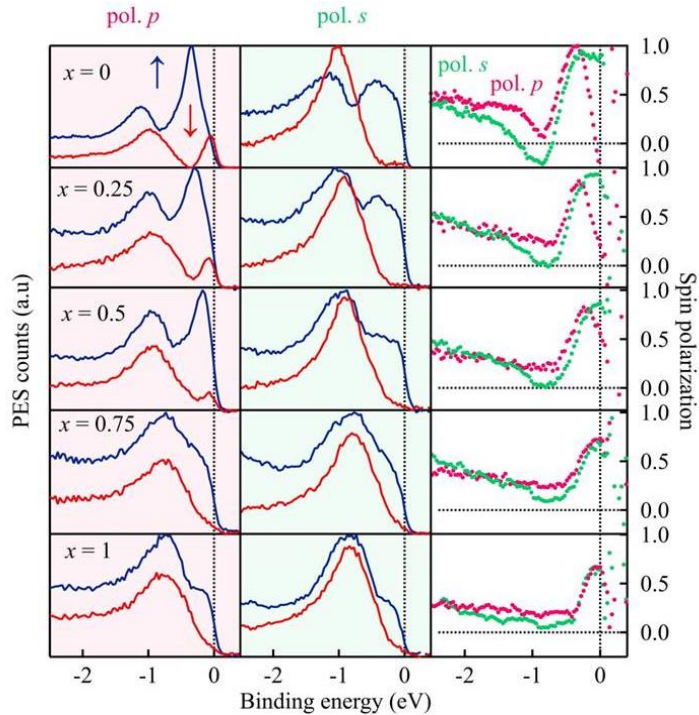

*Figure 2: spin-resolved photoemission spectra using P photon polarization (left), S photon polarization (middle) and resulting spin polarization curves (right) for the Co$_2$MnAl$_x$Si$_{1-x}$ series,*

The radiofrequency magnetic dynamics of the films were thus studied using ferromagnetic resonance (FMR). The magnetic damping coefficient $\alpha$, the effective magnetic moment $M_s$ (close to the true moment in our films due to very small anisotropy – see [22]), and the inhomogeneous linewidth $\Delta f_0$ were thus extracted from the measurements performed on the $Co_2MnAl_xSi_{1-x}$(001) series. The results obtained on the same series used for photoemission experiments are shown in table I. As shown in figure 3, a clear correlation is observed between the spin polarization at $E_F$ and the magnetic damping coefficient $\alpha$, as theoretically expected. An ultra-low $\alpha$ value was obtained for $Co_2MnSi$ (x=0) due to the large spin gap [22]. By substituting Al by Si, the magnetic damping increase is explained by the decrease of the spin polarization.

| $Co_2MnAl_xSi_{1-x}$ | Spin polarization (%) | $M_s$ ($\mu_B$/f.u.) | $\alpha$ (x $10^{-3}$) | $\Delta f_0$ (MHz) | $g$ factor (±0.01) |
|---|---|---|---|---|---|
| x = 0 | 97±3 | 5.08 | 0.46±0.05 | 14.3 | 2.01 |
| x = 0.25 | 90±3 | 4.85 | 0.73±0.15 | 21.7 | 1.99 |
| x = 0.5 | 83±3 | 4.85 | 0.68±0.15 | 9 | 2.01 |
| x = 0.75 | 70±3 | 4.8 | 1.00±0.05 | 81.5 | 2.00 |
| x = 1 | 63±3 | 4.32 | 1.10±0.05 | 22 | 2.01 |

*Table 1: data extracted from spin-resolved photoemission and ferromagnetic resonance experiments performed on the $Co_2MnAl_xSi_{1-x}$ series.*

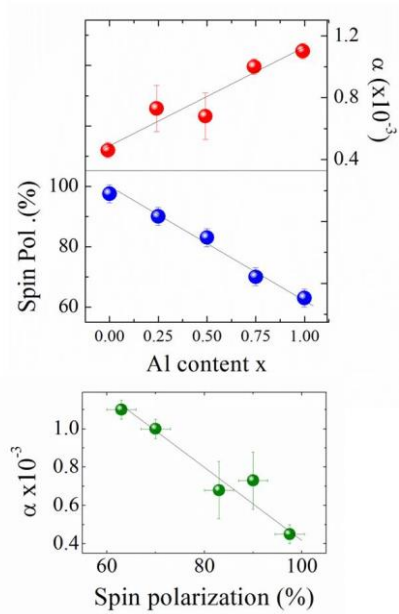

*Figure 3: -top- spin polarization and magnetic damping dependence with Al content for the $Co_2MnAl_xSi_{1-x}$ series and –bottom- magnetic damping versus spin polarization. The lines are guide to the eyes.*

In addition, the magnetization is also observed to decrease with *x* in agreement with the Slater-Pauling description of the valence band electrons in Heusler compounds [25]. Indeed, as a 5 µ$_B$ magnetic moment per cell is expected for Co$_2$MnSi (type IV valence electrons), it should decrease to 4 when replacing Si by Al (type III) as actually observed (Table I). Finally, the FMR susceptibilities reach extremely small inhomogeneous linewidth *Δf$_0$,* a proof of the excellent homogeneity of the magnetic properties (hence a high crystal quality) in our films.

Figure 4(a) shows the ultrafast demagnetization curves measured on the same Co$_2$MnAl$_x$Si$_{1-x}$ series with a maximum magnetization quenching ~15%. The temporal changes of the Kerr signals $\Delta\theta_k(t)$ were normalized by the saturation value $\theta_k$ just before the pump laser excitation. The time evolution of magnetization on sub-picosecond timescales can be fitted according to Eq. (2) in terms of the three-Temperature Model (3TM) [26], which describes the energy distribution among electrons, phonons, and spins after laser excitation.

$$-\frac{\Delta M(t)}{M} = \left\{\left[\frac{A_1}{(t/\tau_0+1)^{0.5}} - \frac{A_2\tau_E - A_1\tau_M}{\tau_E - \tau_M}e^{-\frac{t}{\tau_M}} - \frac{\tau_E(A_1 - A_2)}{\tau_E - \tau_M}e^{-\frac{t}{\tau_E}}\right]\Theta(t)\right\} * G(t,\tau_G) \quad (2)$$

where $G(t,\tau_G)$ represents the convolution product with the Gaussian laser pulse profile, $\tau_G$ is the full width at half maximum (FWHM) of the laser pulses. $\Theta(t)$ is the Heavyside function. The constant A$_1$ represents the amplitude of demagnetization obtained after equilibrium between the electrons, spins, and phonons is reestablished while A$_2$ is proportional to the initial electron temperature raise. The two critical time parameters $\tau_M, \tau_E$ are the ultrafast demagnetization time and magnetization recovery time, respectively. In the low fluence regime, which corresponds to our measurements, $\tau_E$ becomes close to the electron-phonon relaxation time. A unique value of $\tau_E = 550 \pm 20\, fs$ was used for fitting the demagnetization curves for all samples. The ultrafast demagnetization time $\tau_M$ decreases from $380 \pm 10$ fs for Co$_2$MnSi to $165 \pm 10$ fs for Co$_2$MnAl (Figure 4b). The evolution of the demagnetization time with both spin polarization P and Gilbert damping $\alpha$ is presented in figure 4c and 4d. A clear linear variation between $1/\tau_M$ and $1 - P$ is observed in this series. As the magnetic damping $\alpha$ is proportional to P here, this means that $1/\tau_M$ is proportional to $\alpha$ too. A similar relation between these two parameters was proposed by Koopmans *et al.* [10]. However, they also predicted an influence of the Curie temperature. As the Curie temperature in Heusler compounds changes with the number of valence electrons and because the $Co_2MnAl_xSi_{1-x}$ behave as solid solutions as indicated by the lattice spacing variation (Figure 1b), we thus consider a linear decrease of $T_c$ with x going from 985 K to 697 K as experimentally measured for x=0 and x=1, respectively. To test this possible influence of the Curie temperature on the ultrafast magnetization dynamics, we plot in figure 4d first the product $\tau_M.\alpha$ and second the product $\tau_M.\alpha.T_c(x)/T_c(Co_2MnSi)$. These results demonstrate that the Curie temperature does not influence the ultrafast demagnetization in our samples.

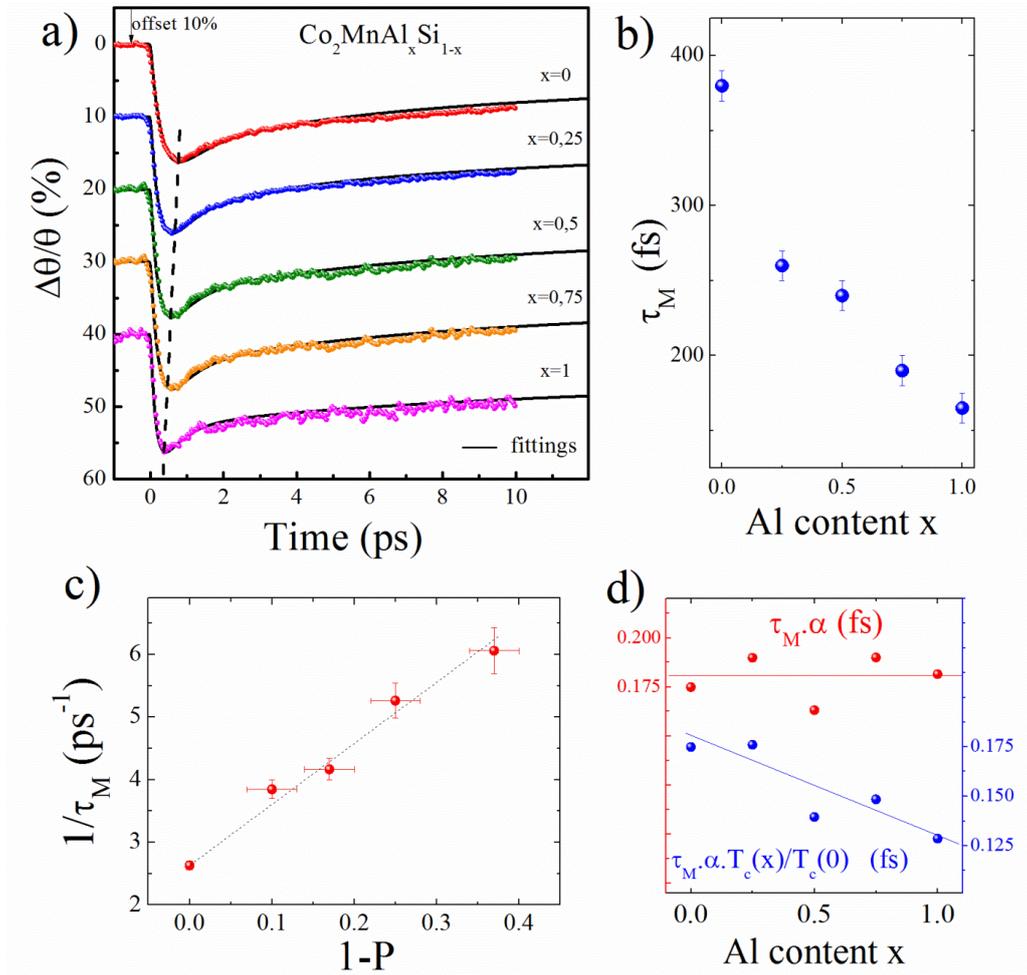

*Figure 4: (a) Ultrafast demagnetization curves obtained for different Al concentration x. The curves have been shifted vertically for sake of clarity. The solid lines represent fitted curves obtained using Eq. (2). (b) Ultrafast demagnetization time as a function of Al content x, (c) the inverse of $\tau_M$ as a function of 1-P, P being the spin polarization at $E_F$, and d) test of Koopmans model with and without taking into account the Curie temperature of the films (see text).*

One can now compare our experimental results with existing theoretical models. We first discuss the dependence between the magnetic damping and the spin polarization. Ultra-low magnetic damping values are predicted in Half-Metal Magnet (HMM) Heusler compounds and explained by the lack of density of state at the Fermi energy for minority spin, or in other words by the full spin polarization [18,27,28]. Consequently, the magnetic damping is expected to increase when creating some states in the minority band structure around the Fermi energy that is when decreasing the spin polarization [28]. If we confirmed in previous experimental works that ultra-low magnetic damping coefficients are actually observed especially on HMM $Co_2MnSi$ and $Co_2MnGe$ [19,22-23], we could not state any quantitative dependence between the damping values and the spin polarization. As prospected, the $Co_2MnAl_xSi_{1-x}$ alloys are shown here to be ideal candidates to address this point. This allows us getting a clear experimental demonstration of these theoretical expectations. Furthermore, a linear dependence between the magnetic damping and the spin polarization is obtained. This behavior may be explained by the mixing of both $L2_1$ and B2 phases in the films. To the best of our knowledge, this experimental result is the first quantitative demonstration of the link between the magnetic damping and spin polarization.

Second, the dependence between the magnetic damping and the demagnetization time observed here is a clear opportunity to test the different theoretical explanations proposed in the literature to explain ultrafast dynamics. In the last 15 years, the influence of the damping on the ultrafast dynamics has been explored, both theoretically and experimentally. The first type of prediction we want to address is the link between the demagnetization time and the electronic structure via the spin polarization P. Using a basic approach considering the Fermi golden rule, several groups [12,13] proposed that the demagnetization process is linked to the population of minority and majority spin states at $E_F$, leading to a dependence of the spin-scattering rate proportional to 1-P [13]. As this spin scattering rate is linked to the inverse of the demagnetization delay time, the $\tau_M \sim (1-P)^{-1}$ law was proposed. This law is clearly verified in our samples series. One should note that this is a strong experimental demonstration since we compare samples grown in the same conditions, so with the same control of the stoichiometry and structural properties.

However, one point is still not clear since much larger demagnetization times in the picosecond timescale would be expected for large band gap and full spin-polarization. In the case of small band gap of the order of 0.1 eV, Mann *et al* [13] showed that thermal effects from the heated electron system lead to a decrease of $\tau_M$. They calculated a reduction of the spin-flip suppression factor from $10^4$ for a gap of 1 eV to 40 for a gap of 0.3 eV. However, the band gap of our Co$_2$MnSi was calculated to be around 0.8 eV with a Fermi energy in the middle of the gap [27,28]. This was corroborated by direct measurement using SR-PES [19, 22]. Therefore, according to their model, we should expect a much longer demagnetization time for Co$_2$MnSi. However, the largest values reported by several groups [13, 29] all on HMM materials are of the same order of magnitude, i.e. around 350 to 400fs. This probably means that a limitation exists due to another physical reason. One hypothesis should be to consider the 1.5eV photon energy which is much larger than the spin gap. During the excitation, the electrons occupying the top minority spin valence band can be directly excited into the conduction band. In a similar way, majority spin electrons are excited at energies higher than the spin band gap. Both of these effects may allow for spin flips scattering and only the majority electrons excited within the spin band gap energy range cannot flip their spins. Even if such photon energy influence is not considered based on the argument that the timescale for photon absorption followed by electronic relaxation is very fast compared to the magnetic relaxation process [16], performing experiments by changing the excitation wavelength to energies below the spin band gap would be very interesting to better understand ultrafast magnetization dynamics.

Concerning the dependence between the demagnetization time and the magnetic damping, different theoretical models have been proposed and two opposite trends were obtained; $\alpha$ and $\tau_M$ being either directly [15] or inversely [10] proportional. From the experimental side, the inverse proportionality between $\tau_M$ and $\alpha$ proposed by Koopmans *et al.* [10] could not be reproduced by doping a thin Permalloy film with rare-earth atoms [14]. However, the introduction of these rare-earth elements strongly modifies the magnetic relaxation properties and could induce different relaxation channels for $\tau_M$ and $\alpha$ [30]. Zhang *et al.* performed a similar study using thin Co/Ni multilayers and observed a direct proportionality between $\tau_M$ and $\alpha$ [15]. However, the damping extracted in their study should be strongly influenced by the heavy metal Pt capping and seed layers which may induce strong spin pumping effect during the magnetization precession [30]. Furthermore, they did not take into account the influence of the Curie temperature. Therefore, in these studies, extrinsic effects might influence the magnetization dynamics in a different way on both time scales which makes more complex the comparison between theory and experiments. Therefore, our results offer a nice opportunity to disentangle these different effects. According to different studies, the ultrafast demagnetization slows down when approaching the Curie temperature [10,16,32,33]. In other words, a larger difference between the initial temperature and $T_c$ would lead to a faster demagnetization. In our samples, $T_c$ goes up from Co$_2$MnAl to Co$_2$MnSi, whereas the demagnetization process becomes slower. Therefore, we conclude that, in the present case, the Curie temperatures of our samples are too high to affect $\tau_M$ which only depends on the intrinsic properties of the films, i.e. Gilbert damping and spin polarization. This also clarifies some points reported by Müller *et al.* work [12]. In their paper, they first reported a very fast demagnetization process in Co$_2$MnSi(110) and second a slow one in CrO$_2$ and LaSrMnO$_3$ films with $T_c$ values close to room temperature (390 K 360 K respectively). Therefore, it is not possible to state whether the very slow demagnetization process in these compounds is due to a low $T_c$ or a large spin polarization. Furthermore, recent experimental results demonstrated a large decrease in the spin polarization at the Fermi level in CrO$_2$ as function of the temperature, resulting in less than 50% at 300 K [34]. In our samples we disentangle these two effects and the longest demagnetization time is found for Co$_2$MnSi ($\tau_M = 380 fs$), a true half-metal magnet with a 0.8 eV spin gap and a large $T_c$.

In summary, we first demonstrate experimentally that substituting Si by Al in Co$_2$MnAl$_x$Si$_{1-x}$ Heusler compounds allows us to get a tunable spin polarization at E$_F$ from ~60% in Co$_2$MnAl to 100% in Co$_2$MnSi, indicating the transition from metallic to half metallic behaviors. Second, a strong correlation between the spin polarization and the Gilbert magnetic damping is established in these films. This confirms the theoretical justification of ultra-low magnetic damping in Half-Metal-Magnets as a consequence of the spin gap. Third, the ultrafast spin dynamics results also nicely confirm that the spin gap is at the origin of the increase of the relaxation time. Our experiments allow us to go further by establishing clear relationships between the spin polarization, the magnetic damping and the demagnetization time. An inverse relationship between demagnetization time and Gilbert damping is established in these alloys, which agrees well with the model proposed by Mann *et al.* [13] and with Koopmans *et al.* [10] but without considering any influence of Curie temperature much larger than room temperature in these films.

**Experimental section**

$Co_2MnAl_xSi_{1-x}$(001) quaternary Heusler compounds are grown by Molecular Beam Epitaxy using an MBE machine equipped with 24 materials. The stoichiometry is accurately controlled during the growth by calibration of the Co, Mn, Si and Al atomic fluxes using a quartz microbalance located at the place of the sample. The error on each element concentration is less than 1% [23]. The films are grown directly on MgO(001) substrates, with the epitaxial relationship [100] (001) MgO // [110] (001) Heusler compound. The thickness is fixed to 20nm.

The photoemission experiments were done at the CASSIOPEE beamline at SOLEIL synchrotron source. The films were grown in a MBE connected to the beamline (see [19,22,35] for details). The SR-PES spectra were obtained by using the largest slit acceptance of the detector (+/- 8°) at an angle of 8° of the normal axis of the surface. Such geometry allows us to analyze all the reciprocal space on similar polycrystalline films [23].

The radiofrequency magnetic dynamics of the films were thus studied using ferromagnetic resonance (FMR). A Vectorial Network Analyzor FMR set-up was used in the perpendicular geometry (see [22] for experimental details) where the static magnetic field is applied out of the plane of the film in order to avoid extrinsic broadening of the linewidth due to the 2-magnons scattering [36,37].

Ultrafast magnetization dynamics were investigated using polar time-resolved magneto-optical Kerr (TR-MOKE) experiments. An amplified Ti-sapphire laser producing 35 fs pulses at 800 nm with a repetition rate of 5 KHz is used. The pump beam is kept at the fundamental mode and is focused down to spot size of $\sim 260 \ \mu m$ while the probe is frequency doubled to 400 nm and focused to a spot size of $\sim 60 \ \mu m$. Samples were magnetically saturated along the out-of-plane axis by applying a 1T magnetic field.

**Acknowledgement**

This work was supported partly by the French PIA project "Lorraine Université d'Excellence", reference ANR-15-IDEX-04-LUE, and by the Agence Nationale de la Recherche (France) under contract no. ANR-17-CE24-0008 (CHIPMuNCS).